\documentclass[sigplan,nonacm,10pt]{acmart}
\AtBeginDocument{%
  }

\setcopyright{acmlicensed}
\copyrightyear{2018}  
\acmYear{2018}  
\acmDOI{XXXXXXX.XXXXXXX}

\acmConference[Conference acronym 'XX]{Make sure to enter the correct
  conference title from your rights confirmation emai}{June 03--05,
  2018}{Woodstock, NY}  
\acmISBN{978-1-4503-XXXX-X/18/06}

\begin{document}

\title{Data Imputation using Large Language Model to Accelerate Recommender System}

\author{Zhicheng Ding}
\email{zhicheng.ding@columbia.edu}
\affiliation{%
  \institution{Columbia University}
  \city{New York}
  \state{NY}
  \country{USA}
}

\author{Jiahao Tian}
\email{jtian83@gatech.edu}
\affiliation{%
  \institution{Georgia Institute of Technology}
  \city{Atlanta}
  \state{Georgia}
  \country{USA}
}

\author{Zhenkai Wang}
\email{kay.zhenkai.wang@utexas.edu}
\affiliation{%
  \institution{The University of Texas at Austin}
  \city{Austin}
  \state{Texas}
  \country{USA}
}

\author{Jinman Zhao}
\email{jinman.zhao@mail.utoronto.ca}
\affiliation{%
  \institution{Univeristy of Toronto}
  \city{Toronto}
  \state{Ontario}
  \country{Canada}
}

\author{Siyang Li}
\email{lisiyang98@hotmail.com}
\affiliation{%
  \institution{Pace University}
  \city{New York}
  \state{NY}
  \country{USA}
}


\renewcommand{\shortauthors}{Zhicheng et al.}

\begin{abstract}
The importance of recommender systems continues to grow as the volume of data generated increases. A robust recommender system can significantly enhance user experience and engagement. However, these systems often face challenges due to missing data, which can arise from various factors, including user privacy concerns and other reasons. In this paper, we propose a framework to address the challenge of sparse and missing data in recommendation systems, a significant hurdle in the age of big data. Traditional imputation methods struggle to capture complex relationships within the data. We propose a novel approach that uses fine-tuned Large Language Model (LLM) to impute missing values for recommendation tasks. LLM, trained on vast amounts of text, is able to understand complex relationships among data and intelligently fill in missing information. We evaluate our LLM-based imputation method across various tasks within the recommendation system domain, including single classification, multi-classification, and regression compared to classical data imputation methods. By demonstrating the superiority of LLM imputation over traditional methods, we establish its potential for improving recommendation system performance.

\end{abstract}



\keywords{Large Language Model, Data Imputation, Recommender System}


\maketitle

\section{Introduction}
\label{sec:intro}
The exponential growth of big data has revolutionized many fields, offering unprecedented access to vast amounts of information. Researchers can find tons of information for uncovering patterns and making informed decisions~\cite{YAO2024100211, Belyaeva2023}. However, this abundance often masks a hidden adversary: sparse and small data. Missing information, often due to user inactivity, limited data collection, or technical constraints, can significantly hinder the effectiveness of big data models~\cite{Fazlikhani2018}. This is particularly true in recommendation systems, where personalized experiences hinge on a rich understanding of users and items, incomplete data significantly hinders the ability to generate accurate suggestions~\cite{Acharya2023}. Traditional statistical methods for data imputation, like mean or median imputation, often fall short in capturing the complex relationships and underlying context within the data~\cite{jin2024apeer, jin2024learning}.

This paper tackles this challenge by proposing a novel approach that leverages the transformative power of LLM to address the challenge of data imputation in recommendation systems. LLMs, with their remarkable ability to process and understand vast amounts of natural language, possess the potential to intelligently fill in these missing data points. By harnessing the LLM's capability to learn intricate relationships and context from large text corpora, our proposed method aims to impute data that is not only statistically sound but also semantically meaningful~\cite{Jäger2021}. This enriched data can then be utilized by recommendation systems to generate more accurate and personalized suggestions for users.

Focusing on the domain of recommendation systems, we explore the specific application of LLM-based data imputation. Recommender systems rely heavily on comprehensive user and item data to generate personalized suggestions that resonate with individual preferences. By effectively imputing missing values, we aim to create a more complete picture of user behavior and item characteristics. This, in turn, allows the recommendation system to generate more accurate and relevant suggestions, ultimately enhancing the user experience.

We meticulously design a series of experiments to evaluate the effectiveness of our approach. The experiences encompass a diverse range of classification and regression tasks. These experiments delve into single classification, where the system predicts a single category for an item, multi-classification, which allows for assigning multiple categories, and regression, where the focus is on predicting continuous values like ratings. By demonstrating the superiority of LLM imputation over traditional statistical methods across these varied scenarios, we aim to establish its significance as a game-changer in improving the performance of recommendation systems.

To comprehensively assess the effectiveness of LLM-based imputation, we conduct rigorous experiments across a diverse range of tasks within the recommender system domain. These experiments encompass single classification, where the system predicts a single category for an item (e.g., AD recommendation), multi-classification, where multiple categories can be assigned (e.g., multiple categorical movies recommendation), and regression, which focuses on predicting continuous values like ratings or purchase likelihood (e.g., movie rating prediction). By demonstrating the advantage of LLM data imputation over traditional statistical methods in these varied scenarios, we experiment our proposed approach in different recommendations system tasks with different datasets. In summary, our paper makes the following primary contributions:
\begin{itemize}
    \item [$\bullet$] We propose a novel approach that utilize LLM to impute missing data which aims to handle data sparsity and data bias issue.
    
    \item [$\bullet$] We further utilize the imputed data and evaluate in the recommendation system which shows improvement to other statistical data imputation strategy.

    \item [$\bullet$] Extensive experiment are done to further prove that LLM-based data imputation works better in single classification, multiple classification task and regression recommendation tasks.

\end{itemize}

\section{Related Work}
\subsection{Data Imputation}
Data imputation has been studied extensively in both statistics and machine learning, with a rich history of methodological development. Traditional methods, such as replacing missing values with constants (e.g., zero, minimum, maximum) or aggregated measures (mean, median, most frequent), are simple but often introduce bias into the dataset~\cite{newman2014missing}. To mitigate this limitation, more sophisticated techniques have been developed, including k-Nearest Neighbors (kNN) imputation, which imputes missing values based on similar data points, and model-based methods that leverage statistical models to predict missing values~\cite{ijgi9040227, PengSVM}. Recently, research has focused more on machine learning algorithms for imputation, such as matrix factorization and deep learning techniques \cite{hwang2018data}
, which can handle complex patterns and relationships within the data for more accurate imputations. However, choosing the optimal imputation method remains a critical task, which is influenced by factors such as data type, missing data mechanism (missing completely at random, missing at random, or missing not at random), the amount and pattern of missing data, and specific analytical goals~\cite{Ben2023}.

\subsection{Large Language Model}

LLM is trained on massive amounts of text data, have shown promise due to their ability to capture complex relationships and semantic information within data. This capability allows them to potentially impute missing values in a more reliable way than traditional methods. For instance, some approaches treat imputation as a classification task, where the LLM predicts the most likely value for the missing entry based on the surrounding data~\cite{XinjinVehicle,XinjinVehicleDL}. Others leverage the generative nature of LLMs to create a distribution of possible values, providing a more comprehensive picture of the imputation uncertainty. While promising, research on LLM-based data imputation is still evolving. Areas of exploration include mitigating potential biases present in training data and ensuring the imputed values maintain data integrity, particularly in sensitive domains like healthcare~\cite{wu2024nextgptanytoanymultimodalllm,deng2024composerx}. Overall, LLMs offer a new avenue for tackling missing data issues, with the potential to improve the accuracy and robustness of data analysis in various fields. There are also many successful LLM applications such as in Relation Extration(RE)~\cite{gptre}, NER~\cite{gptner,zeroshotner}, feature engineering~\cite{gptsignal}, text summarization~\cite{goyal2023news}  and sentiment analysis~\cite{sun2023sentiment}.

\subsection{Recommender System}
Recommender System(RS) is used to generate meaningful suggestions to a collection of users for items or products that might interest them. RS can be divided into personalized~\cite{wu2022knowledge,zheng2022explainable,zhao2022two} and group-based~\cite{stratigi2022sequential,kumar2022automatically,zan2021uda,sato2022enumerating,zhang2022gbert} systems. In recent years, neural networks like CNN~\cite{an2022design}, GCN~\cite{kipf2016semi}, GraphSAGE~\cite{hamilton2017inductive}, and others have significantly enhanced RS models.

Data sparsity is a persistent challenge in such systems, significantly impacting the accuracy and effectiveness of recommendations. Collaborative filtering techniques, a mainstay in recommendation systems, struggle when user-item interaction matrices are highly sparse, with many missing entries. This sparsity makes it difficult to identify similar users or items for accurate recommendations~\cite{Lubos2024}. More and more research has explored various approaches to address this issue, focus on developing robust recommendation systems that can effectively handle data sparsity and deliver personalized recommendations even with limited user-item interactions. In this paper, we aims to handle those missing data using LLM-based data imputation technology.

\section{Method}
\label{sec:method}

\begin{figure*}
  \centering
  \includegraphics[width=\linewidth]{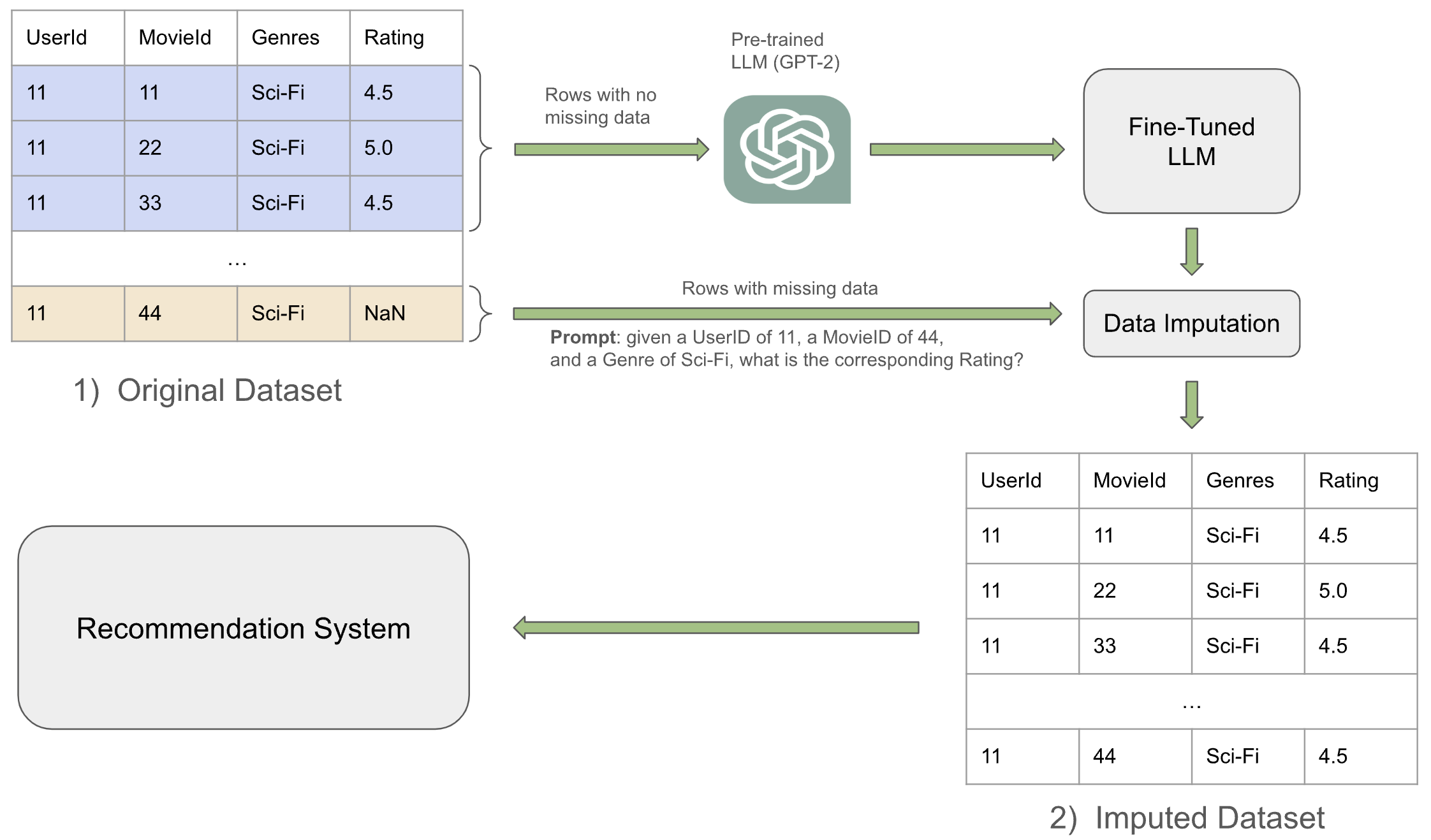}
  \caption{Framework of our proposed method. Original dataset contain missing data. Using complete data to fine-tune LLM which can be further utilized to impute the missing data. After that, complete tabular data are used to feed into Recommender System.}
  \label{fig:framework}
\end{figure*}
In this section, we present the major components of our proposed architecture. We start with fine-tuning a pre-trained model using our task-specific dataset that only contains complete data. This fine-tuned model is then employed to impute missing data. The resulting dataset, which contains both the complete and imputed data, is then fed into the recommender system. Figure \ref{fig:framework} provides a visual representation of the architecture. Detailed discussions are provided in the subsequent sections.

\subsection{Data Preparation}
To tailor LLM for our specific task and data at hand, 
we first need to fine-tune LLM. By fine-tuning a model on a much smaller dataset, its performance on the task can be improved while preserving its general language knowledge. We divide our dataset into two subsets, one is the set that only contains complete data, and the other contains data entries with missing values.

\subsection{Fine-tune LLM Model}
For the fine-tuning process, we utilize the complete dataset to enable the model to learn task-specific information. We adopt Low-Rank Adaptation (LoRA) technique ~\cite{hu2022lora} to achieve efficient fine-tuning of LLM. LLM is typically trained with billions of parameters, rendering comprehensive fine-tuning computationally expensive. LoRA offers a cost-effective alternative by freezing the pre-trained model weights and introducing a set of trainable low-rank adapter parameters. This approach significantly reduces the computational burden associated with fine-tuning while enabling the LLM to adapt to the specific task or domain~\cite{borisov2023language}.

The data flow in the fine-tuning process begins with the collection and pre-processing of task-specific data, which is tokenized and converted into input tensors compatible with the LLM architecture. These tensors are then fed into the pre-trained LLM model. Instead of updating the entire weight matrices, LoRA introduces low-rank matrices that approximate the necessary updates. During each forward pass, the input data propagates through the attention and feed-forward layers, where the low-rank matrices are applied to modify the output dynamically. The resulting predictions are compared with the ground truth to compute the loss, which is then backpropagated through the model. Only the parameters associated with the low-rank matrices are updated, leaving the original pre-trained weights largely intact. This selective adaptation allows the model to learn task-specific features efficiently while preserving its general language understanding capabilities learned during LLM's original training. The low-rank matrices focus on the most influential components of the dataset, enhancing the model's ability to predict and fill in missing values accurately. This approach not only speeds up the fine-tuning process but also reduces memory and storage requirements,  improving LLM's accuracy on data imputation tasks.

By fine-tuning the pre-trained model with a dataset containing only complete entries, we obtain a LLM that not only retains knowledge from its extensive pre-training but incorporates specific patterns from the current dataset. This approach leverages the model's broad understanding while adapting it to the nuances of our specific task.

\subsection{Data Imputation}
Subsequently, the fine-tuned LLM mentioned above is used to impute missing data. We incorporate existing data information as relevant knowledge into the prompt. Prompts constructed in this manner contain example-specific information and LLM is used to model the distribution of the missing attributes. Note that the prompt can also be constructed to impute multiple values for a single example simultaneously.For instance, given a data entry with attributes \textit{UserId=11}, \textit{MovieId=44}, \textit{Genres=Sci-FI}, and \textit{Rating=NaN} (indicating missing value), the prompt would be formulated as: \textit{"given a UserID of 11, a MovieID of 44, and a Genre of Sci-Fi, what is the corresponding Rating?"}. As a result, LLM will generate the most probable values based on patterns learnt from the training data and the given prompt. Then $NaN$s  are replaced with LLM imputed values. The imputed data is combined with the complete data to form a whole dataset used for training the Recommender System.

\subsection{Evaluation in Recommender System}
To comprehensively assess the efficacy of the LLM-based data imputation approach, the newly constructed dataset was subsequently employed to train a deep-learning-based recommendation system. To achieve a holistic evaluation of the advantages offered by LLM-based imputation, performance metrics was utilized across various task categories, encompassing single classification, multi-class classification, and regression. Within the single classification domain, precision, recall, and F1-score were adopted as the evaluation metrics. For multi-class classification tasks, Recall at k (denoted as R@k) and Normalized Discounted Cumulative Gain at k (denoted as N@k) were employed. Finally, Mean Absolute Error (MAE), Mean Squared Error (MSE), or Root Mean Squared Error (RMSE) were leveraged to assess the performance of the regression task.

\begin{table*}[h]
    \centering
    \begin{tabular}{l|c c c | c c c}
         model & R@3$\uparrow$ & R@5$\uparrow$ & R@10$\uparrow$ & N@3$\uparrow$ & N@5$\uparrow$ & N@10$\uparrow$\\
         \hline 
         Case-Wise Deletion & 0.2510 & 0.3470 & 0.6050 & 0.4853 & 0.5381 & 0.7011 \\
         Zero & 0.2370 & 0.3250 & 0.5760 & 0.4412 & 0.5005 & 0.6629 \\
         Mean & 0.2610 & 0.3490 & 0.6110 & 0.5064 & 0.5455 & 0.7294 \\
         Knn & 0.2880 & 0.3870 & 0.6420 & 0.5213 & 0.5674 & 0.7331 \\
         Multivariate & 0.2760 & 0.3680 & 0.6440 & 0.5154 & 0.5401 & 0.7420 \\
         LLM & \textbf{0.2930} & \textbf{0.4050} & \textbf{0.6530} & \textbf{0.5692} & \textbf{0.6216} & \textbf{0.7632}
    \end{tabular}
    \caption{Comparison among LLM-based and statistical data imputation on multiple classification task}
    \label{table:multiple-classification}
\end{table*}
\section{Experiment}
\label{sec:experiment}

\subsection{Model and Dataset}
We chose to utilize the pre-trained distilled version of GPT-2 model \cite{sanh2019distilbert}
 as our LLM because of its open-source accessibility and proven effectiveness across a wide range of tasks. For the choice of dataset, we use AdClick~\cite{criteo-display-ad-challenge} and MovieLen~\cite{movie-recommender-system-2022} dataset for this experiment due to the fact that these are large well-structured dataset without requiring extensive data cleaning. In addition, there have been many researches conducted on these two datasets and they are available for both classification and regression tasks. 

The original dataset is a structured tabular dataset. To simulate a dataset with missing values, we introduce missing data in a controlled manner. For each column in the dataset, we randomly select 5\% of the data points and mark them as missing.
This selection is done independently for each column so that the rows with missing data will vary from column to column. Due to the independent selection process, more than 5\% of the rows may contain at least one missing value. Here we break down into 3 different recommendation tasks:

\begin{itemize}
    \item \textbf{Single Classification:} We leverage the AD Click dataset to evaluate the effectiveness of our proposed architecture. The imputed data is then fed into a recommendation system designed to classify user clicks on advertisements. This approach aims to improve the accuracy of predicting user engagement with targeted advertising.
    \item \textbf{Multiple Classification: }  We employ the widely used MovieLens dataset to assess the impact of LLM-based data imputation on movie recommendations. The imputed data is subsequently utilized by a recommendation system to suggest a personalized list of top-k movies for each user. This research aims to enhance the effectiveness of recommendation systems by addressing data sparsity issues.
    \item \textbf{Regression: } Building upon the MovieLens dataset, we investigate the use of LLMs for data imputation in predicting user ratings. The imputed data is then incorporated into a recommendation system tasked with predicting user ratings on a scale of 0.0 to 5.0. This approach seeks to improve the accuracy of rating predictions within recommendation systems. 
\end{itemize}     

\subsection{Baselines}
The pre-processed datasets above contain about 5\% rows with missing data. To evaluate the effectiveness and efficacy of the proposed LLM-based data imputation, we compare its performance against the following competing baseline methods: 
\begin{itemize}
    \item \textbf{Case-Wise Deletion: } without imputing missing data, feed data directly to the recommender system and discard examples with one or more missing values.
    \item \textbf{Zero Imputation: } replaces all missing numeric values with 0.
    \item \textbf{Mean Imputation: } calculates the arithmetic mean of the column and replaces missing values with it.
    \item \textbf{KNN Imputation: } imputes missing values using k-Nearest Neighbors. Each sample’s missing values are imputed using the mean value from n-th neighbors nearest neighbors found in the training set.
    \item \textbf{Multivariate Imputation: } estimates each feature from all the others and imputes missing values by modeling each feature with missing values as a function of other features in a round-robin fashion~\cite{scikit-learn}.
\end{itemize}

\subsection{Evaluation}
To better evaluate LLM-based data imputation technique, we evaluate our model's performance across three tasks: single classification, multi-class classification, and regression. Two benchmark datasets, AD click and MovieLens, are used. For both datasets, we meticulously curate the data to achieve a targeted missing value ratio of approximately 5\%. Then, we feed the data with no imputation into our recommendation system for baseline performance. The DLRM~\cite{DLRM19} model is utilized for this purpose and we randomly split of 60/20/20 is used for training, testing, and validation, respectively.
In addition, we applied statistical methods (mean, zero, KNN, and iterative) and our LLM-based approaches. The imputed data by different approaches will feed into DLRM one by one following the same 60/20/20 ratio for the evaluation.

The detailed results of single classification task are presented in Table~\ref{table:single-classification}, with the top and second-highest performing models highlighted for clarity. While LLM-based data imputation approach did not achieve the absolute top performance in this particular task, as we will demonstrate in the following section, it exhibits potential for superior performance in more complex scenarios. Table~\ref{table:multiple-classification} presents the results of multiple classification task. Due to the richer metadata and intricate relationships within the MovieLens dataset, the LLM-based model demonstrates a clear advantage over other models. Finally, we evaluate the effectiveness of LLM-based data imputation within a regression task comparing with statistical methods. 
Table~\ref{table:regression} showcases the results, highlighting the superior performance of the LLM-based data imputation approach compared to other models.

\begin{table}[]
    \centering
    \begin{tabular}{l|c c c}
         model & precision$\uparrow$ & recall$\uparrow$ & f1-score$\uparrow$\\
         \hline 
         Case-Wise Deletion & 0.1980 & 0.4450 & 0.2740 \\
         Zero & 0.7207 & 0.7200& 0.7200 \\
         Mean & 0.8846 & 0.8700 & 0.8702 \\
         Knn & \textbf{0.9192} & \textbf{0.9150} & \textbf{0.9150} \\
         Multivariate & 0.8970 & 0.8900 & 0.8903 \\
         LLM & \underline{0.9071} & \underline{0.9001} & \underline{0.9003}
    \end{tabular}
    \caption{Comparison among LLM-based and statistical data imputation on single classification task}
    \label{table:single-classification}
\end{table}


Finally, we evaluate the effectiveness of LLM-based data imputation within a regression task comparing with statistical methods. 
Table~\ref{table:regression} showcases the results, highlighting the superior performance of the LLM-based data imputation approach compared to other models.

\begin{table}[]
    \centering
    \begin{tabular}{l|c c c}
         model & MAE$\downarrow$ & MSE$\downarrow$ & RMSE$\downarrow$ \\
         \hline 
         Case-Wise Deletion & 0.7659 & 0.9792 & 0.9895 \\
         Zero & 0.7798 & 0.9928 & 0.9964 \\
         Mean & 0.7804 & 0.9883 & 0.9942 \\
         Knn & 0.7791 & 0.9909 & 0.9955 \\
         Multivariate & 0.7785 & 0.9887 & 0.9943 \\
         LLM & \textbf{0.7612} & \textbf{0.9647} & \textbf{0.9822}
    \end{tabular}
    \caption{Comparison among LLM-based and statistical data imputation on regression task}
    \label{table:regression}
\end{table}


\section{Conclusion}
\label{sec:conclusion}

In conclusion, this paper proposes a novel approach that leverages the power of LLM to address missing data in the Recommender System. By imputing missing data in a semantically meaningful way, our method enriches data and allows the Recommender System to generate more accurate and personalized suggestions, ultimately enhancing user experience. We extensively evaluate our approach across various recommender system tasks, demonstrating its effectiveness in improving performance compared to traditional data imputation methods. The implications of this research extend beyond recommender systems, opening new avenues for utilizing LLMs to mitigate data sparsity and small sample size issues in big data models, leading to a more robust Recommender System.

\bibliographystyle{ACM-Reference-Format}
\bibliography{main}

\end{document}